\documentclass[final]{IEEEtran}
\newlength \figwidth
\usepackage{cite}
\ifCLASSINFOpdf
  \usepackage[pdftex]{graphicx}
	\usepackage{epstopdf}
  \graphicspath{{../Figures/}}
  \DeclareGraphicsExtensions{.eps}
\else
  \usepackage[dvips,draft]{graphicx}
\fi
\usepackage[cmex10]{amsmath}
\usepackage{amssymb}
\usepackage{pifont}
\usepackage{amsthm}
\usepackage{url}
\usepackage{dsfont}
\usepackage{tabulary}
\usepackage{multirow}
\usepackage{color}

\usepackage{color}
\usepackage{times}
\usepackage{verbatim}
\usepackage{floatflt}
\usepackage{enumerate}
\usepackage{array}
\usepackage{multicol,afterpage,wrapfig} 
\usepackage{tikz}
\usepackage{algorithm}
\usepackage[font=small]{caption}
\usepackage[noend]{algpseudocode}
\makeatletter
\def\BState{\State\hskip-\ALG@thistlm}
\makeatother
\usepackage{amsmath,amssymb,eucal}
\usepackage[utf8]{inputenc}
\usepackage{epsfig}
\usepackage{exscale}
\usepackage{latexsym}
\usepackage{verbatim}
\usepackage{amsfonts}
\usepackage{multirow}
\usepackage{cite}
\usepackage{balance}
\usepackage{color}
\usepackage{transparent}
\usepackage{import}
\usepackage{mathtools}
\usepackage{tikz}
\usetikzlibrary{automata,arrows,positioning,calc}
\usepackage{bbm}
\usepackage[printonlyused,withpage]{acronym}
\usepackage{tabu}
\usepackage{longtable}
\usepackage{balance}
\usepackage{footnote}
\usepackage{tablefootnote}
\usepackage{subfig}

\def\BibTeX{{\rm B\kern-.05em{\sc i\kern-.025em b}\kern-.08em
    T\kern-.1667em\lower.7ex\hbox{E}\kern-.125emX}}

\newcommand*\xbar[1]{%
  \hbox{%
    \vbox{%
      \hrule height 0.5pt 
      \kern0.36ex
      \hbox{%
        \kern-0.12em
        \ensuremath{#1}%
        \kern-0.12em
      }%
    }%
  }%
}

\newfont{\bbb}{msbm10 scaled 500}

\newfont{\bb}{msbm10 scaled 1100}

\newcommand{\executeiffilenewer}[3]{%
\ifnum\pdfstrcmp{\pdffilemoddate{#1}}%
{\pdffilemoddate{#2}}>0%
{\immediate\write18{#3}}\fi%
}

\allowdisplaybreaks

\IEEEoverridecommandlockouts
\begin{document}
\pagenumbering{gobble}

\newtheorem{Theorem}{\bf Theorem}
\newtheorem{Corollary}{\bf Corollary}
\newtheorem{Remark}{\bf Remark}
\newtheorem{Lemma}{\bf Lemma}
\newtheorem{Proposition}{\bf Proposition}
\newtheorem{Assumption}{\bf Assumption}
\newtheorem{Definition}{\bf Definition}
\title{Massive MIMO Unlicensed:\\A New Approach to Dynamic Spectrum Access}
\author{\IEEEauthorblockN{{Adrian~Garcia-Rodriguez{$^{\dag}$}, Giovanni~Geraci{$^{\dag}$}, Lorenzo~Galati~Giordano{$^{\dag}$}, \\ Andrea~Bonfante{$^{\dag}$}, Ming~Ding{$^{\ddag}$}, and David~L\'{o}pez-P\'{e}rez{$^{\dag}$}}}\\
\normalsize\IEEEauthorblockA{{$^{\dag}$}\emph{Nokia Bell Labs, Dublin, Ireland \\}}
\normalsize\IEEEauthorblockA{{$^{\ddag}$}\emph{Data61, Sydney, Australia}}}

\maketitle
\thispagestyle{empty}
\begin{abstract}
Nowadays, the demand for wireless mobile services is copious, and will continue increasing in the near future. Mobile cellular operators are therefore looking at the unlicensed spectrum as an economical supplement to augment the capacity of their soon-to-be overloaded networks. The same unlicensed bands are luring internet service providers, venue owners, and authorities into autonomously setting up and managing their high-performance private networks. In light of this exciting future, ensuring coexistence between multiple unlicensed technologies becomes a pivotal issue. So far this issue has been merely addressed via inefficient sharing schemes based on intermittent transmission. In this article, we present the fundamentals and the main challenges behind \emph{massive MIMO unlicensed}, a brand-new approach for technology coexistence in the unlicensed bands, which is envisioned to boost spectrum reuse for a plethora of use cases.
\end{abstract}

\IEEEpeerreviewmaketitle
\section*{Introduction}

While mobile network operators (MNOs) used to see investing in unlicensed frequencies as a means to feed competing technologies, 
they embrace it now as a tool to efficiently address the exponential growth of traffic demands. 
Scarce and costly licensed bands can be relieved by offloading best-effort traffic to unlicensed spectrum. 
Conversely, licensed technologies can take over when unlicensed ones happen to fail, 
thus providing enhanced quality of experience and reliability. 
This allows to promptly deal with reduced coverage, increased interference, or even a radar operating in the same band. 
Better still, with a licensed-plus-unlicensed heterogeneous spectrum, 
MNOs can seamlessly offer larger bandwidths, and thus improved performance to their end users, 
e.g., in terms of higher (peak) data rates.

Different viewpoints and new business models arise when attempting to efficiently use the unlicensed spectrum. 
Broadly speaking, one can divide the actors involved in this dispute into the following two main categories.

$\bullet$ \textbf{Cellular operators:} 
Through a closer look, cellular operators can be further classified into two camps: 
those that in addition to cellular own wireless local area networks (WLANs) --- based on IEEE 802.11 (Wi-Fi) --- and those that do not.
Operators of the former type are keen in reusing their large number of installed WLAN access points. 
These operators are pro long term evolution (LTE)-WLAN aggregation (LWA)-like technologies. 
In essence, LWA efficiently realizes licensed-unlicensed spectrum aggregation through WLAN access points 
by using the Third Generation Partnership Project (3GPP) dual connectivity framework as well as an optimized packet data convergence protocol (PDCP) split~\cite{3GPP36360}. 
In contrast, operators that do not own WLANs prefer a native LTE carrier aggregation technology to directly operate the unlicensed spectrum. 
This is due to its easier management and integration with their existing LTE networks. 
Such approach is adopted in LTE unlicensed (LTE-U) and licensed assisted access (LAA)~\cite{3GPP36889, jindal2015lte}. 
Both enable aggregation of licensed and unlicensed component carriers (CCs) at the medium access control (MAC) layer, 
where a licensed CC must always be present as primary CC, the anchor of the carrier aggregation. 

$\bullet$ \textbf{New wireless providers:} 
In their quest to conquer new vertical markets and their associated revenues, 
network equipment vendors and service providers  have created MulteFire (MF), 
yet another LTE-like industrial standard that operates in the unlicensed band~\cite{MulteFireTechnicalPaper}. 
With the critical feature of not requiring a licensed carrier anchor, 
and therefore allowing stand-alone operation in the unlicensed spectrum, 
MF ushers in a new class of wireless providers, populated by an ecosystem of enterprise, industry, and internet of things (IoT) networks.

\bigskip
Given the broad range of new technologies operating in the unlicensed spectrum, 
guaranteeing seamless inter-technology coexistence is essential, 
particularly with the omnipresent WLANs \cite{jindal2015lte}.
For this reason, and to be fair to traditional service providers,
access to the unlicensed band is strictly regulated, 
and compliance to well defined regulatory requirements is imposed according to the geographical area \cite{3GPP-RP-140808}.
LWA guarantees coexistence with WLAN, 
as it uses WLAN access points to operate in the unlicensed spectrum.
LAA and MF ensure a fair coexistence by implementing listen before talk (LBT) operations that resemble very closely those used by WLAN nodes and are in line with the most strict regulatory requirements. 
It is important to note, however, that all of the above approaches are based on discontinuous transmissions, 
and none allows simultaneous usage of the unlicensed spectrum by two technologies with overlapped coverage areas. 
Such over-polite modus operandi may prove particularly suboptimal in densely deployed scenarios, 
preventing the attainment of high data rates.

Massive multiple-input multiple-output unlicensed (mMIMO-U), invented by the authors of this paper, overcomes this hurdle by integrating new features that could be adopted by LAA, MF, or even WLAN \cite{GerGarLop2016}. 
Specifically, mMIMO-U exploits the spatial awareness provided by a large number of antennas, 
and focuses radiated energy to active users, 
while actively suppressing interference towards and from coexisting neighbors. 
Overall, this results in a large spectrum reuse and thus an improved network performance.


\section*{The mMIMO-U Technology}

Massive MIMO (mMIMO) sub-6 GHz is a key component of the fifth generation (5G) wireless systems, 
with the foreseen role of providing a high-capacity umbrella of ubiquitous coverage.
In mMIMO, cellular base stations (BSs) are equipped with a large number of antennas, 
which provide them with many more spatial degrees of freedom (d.o.f.) than the number of user terminals (UTs) to be served per time-frequency resource~\cite{Mar:10}. 
With such availability of d.o.f., 
the effects of uncorrelated noise and small-scale fading vanish, 
and large spectral efficiencies can be reliably achieved in fast-changing propagation environments.

\begin{figure}[!t]
\centering
\includegraphics[width=0.85\columnwidth]{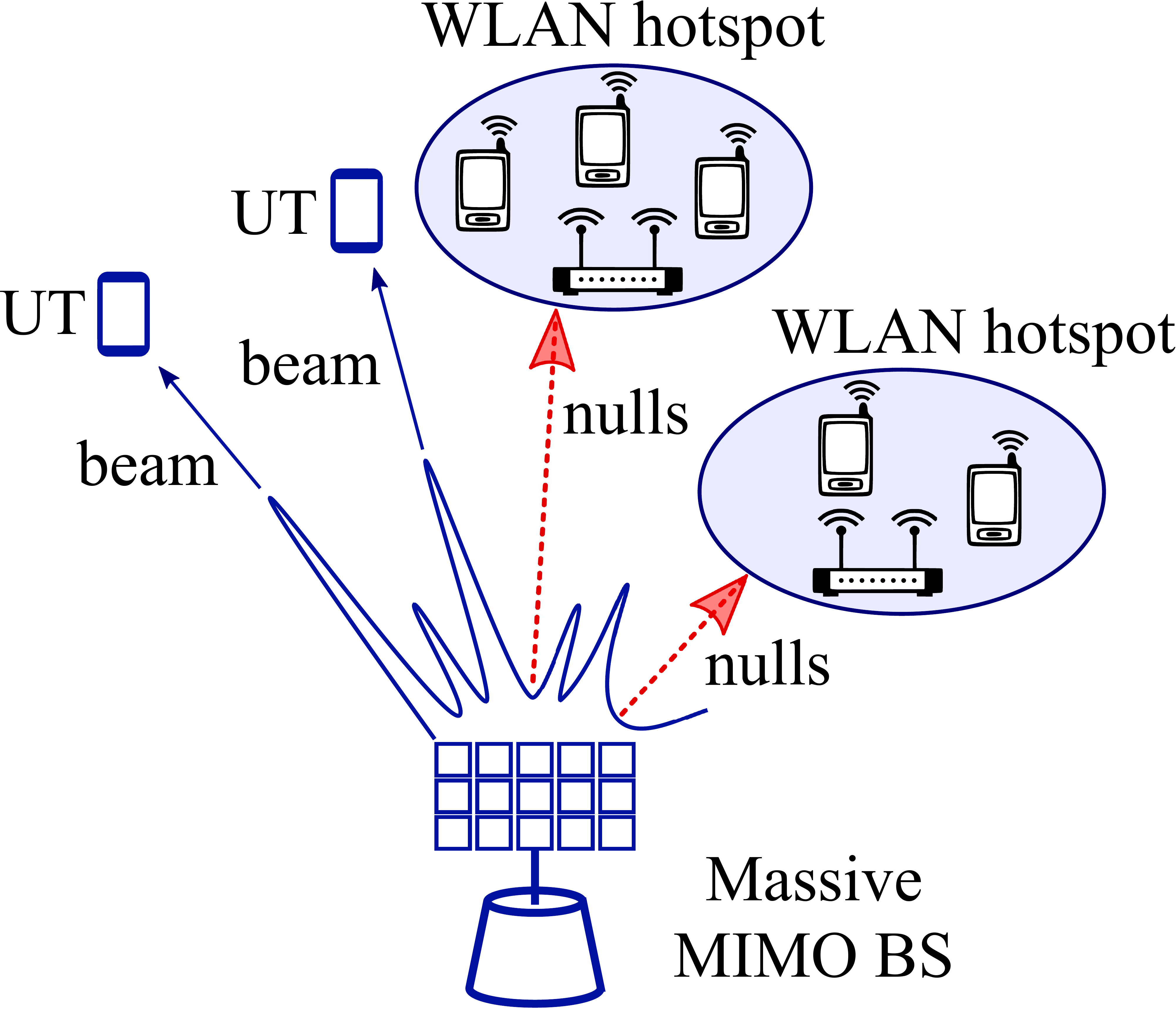}
\caption{Illustration of a mMIMO-U system: A BS multiplexes UTs in the unlicensed band while spatially suppressing interference at neighboring WLAN devices.}
\label{fig:SystemModel}
\end{figure}

In this article, we present a new application of mMIMO, denoted as mMIMO-U, where large-antenna-array BSs operate in unlicensed bands. 
In an attempt to enhance coexistence, 
each BS in mMIMO-U exploits its precise spatial resolution to suppress the mutual interference between itself and other unlicensed terminals sharing the same spectrum, 
e.g., WLAN devices operating in the same coverage area.
In particular, interference is suppressed by placing radiation nulls both during the data transmission phase, 
as illustrated in Fig.~\ref{fig:SystemModel}, 
as well as during the mandatory LBT phase. 
The rationale behind this system design is based on the channel reciprocity of time division duplex (TDD) systems, 
i.e., BSs that do not transmit in a given direction or channel subspace do not need to listen in that direction or channel subspace during the LBT phase either, as a traditional directional antenna would do.

While sacrificing some of the d.o.f. for radiation nulls reduces the mMIMO array gain, 
it provides largely increased opportunities for channel access and spectrum reuse. 
This is the main advantage of mMIMO-U, whose key technical procedures are introduced in the following.

\subsection*{Channel Assessment: Enhanced Listen Before Talk (eLBT)}

In some geographical regions, e.g., Europe and Japan ~\cite{3GPP-RP-140808}, 
each cellular BS must perform LBT prior to data transmission in order to comply with the regulatory requirements. In current technologies, such as LAA and MF, 
a BS must undergo an energy detection procedure before being able to avail of the unlicensed spectrum. 
The energy detection phase consists in evaluating whether the received sum power amounts to less than a regulatory threshold, 
and it lasts for a distributed inter-frame space (DIFS) plus a randomly drawn number of back-off time slots~\cite{PerSta2013}. 
Such approach only allows the transmission of either a single BS or a WLAN device within a certain coverage area, 
which may prove over-conservative and prevent spatial reuse of the same unlicensed spectrum.

In mMIMO-U, the LBT phase is enhanced (eLBT) by placing radiation nulls towards neighboring WLAN devices, 
which include both access points (APs) and stations (STAs). 
In order to place such nulls, 
each BS periodically calculates the dominant eigen-directions of the channel subspace occupied by nearby WLAN devices through a channel covariance estimation procedure \cite{HoyHosTen2014, GerGarLop2016}. 
Throughout the covariance estimation procedure, 
all BSs remain silent and receive a signal which consists of all transmissions from active WLAN devices.\footnote{
When multiple technologies reuse the same unlicensed band, 
the channel covariance estimation includes their transmitted signals, 
and radiation nulls are placed to ensure coexistence among all technologies and WLAN devices.} 
From the WLAN covariance estimate, 
each BS gains spatial awareness and can allocate a certain number of spatial d.o.f. to place radiation nulls towards the dominant WLAN channel eigendirections. 
Let $N_{\rm N}$ be such number, 
and let $N_{\rm A}$ be the number of BS antennas. 
Then, the remaining $N_{\rm A}-N_{\rm N}$ d.o.f. can be used for UT multiplexing. 
Intuitively, the value of $N_{\rm N}$ is chosen by trading beamforming/multiplexing gain at the UTs for enhanced coexistence in the unlicensed spectrum.

During the eLBT phase, 
a BS listens to the transmissions currently taking place in the unlicensed band, 
and it measures the aggregate power of the received signal filtered through the $N_{\rm N}$ radiation nulls. 
Provided that a sufficient number of nulls have been allocated and that these have been well placed, 
the eLBT phase is successful, i.e., no concurrent transmissions are detected. 
A successful eLBT phase allows the BS to access the channel for downlink (DL) transmissions even when one or more WLAN devices are transmitting, 
achieving additional spectrum reuse in the spatial domain. 

\subsection*{Scheduling: WLAN-Aware User Selection}

After a successful eLBT phase, 
cellular BSs and WLAN devices can simultaneously operate in the unlicensed spectrum. 
While the mutual BS-WLAN interference can be suppressed through radiation nulls, 
the same does not hold for the WLAN-to-UT interference. 
Indeed, the latter may degrade the mMIMO-U downlink rates at the targeted UT. 
A WLAN-node-aware user selection process is therefore needed, 
where the mMIMO-U BS may schedule transmissions to UTs in various manners, 
depending on their radio proximity to WLAN devices.

To this end, scheduling metrics that account for the average channel gain between the UT and one or more WLAN APs can be defined. 
In practical implementations, such information can be obtained through the automatic neighbor relations (ANR) function, 
which is already adopted in systems such as LWA~\cite{3GPP36300}. 
When implementing the ANR functionalities, 
BSs are capable of requesting UTs to report measurements containing the received signal strength indicator (RSSI) from a certain WLAN AP. 
The above measurements vary on a large time scale, 
and they can be fed back by the UT to the BS without incurring a significant overhead. 
WLAN RSSI information can be combined with a proportional fair (PF) metric to enforce fairness by accounting for both the UT current and past data rates.

As a result of the WLAN-aware scheduling procedure, 
the mMIMO-U BS can determine the $N_{\rm U}$ UTs that are far from WLAN devices and scheduled for transmission. 
These UTs can thus be served by reusing the same spectrum used by WLAN transmissions, 
as illustrated in Fig.~\ref{fig:SystemModel}.

\begin{figure}[!t]
\centering
\includegraphics[width=0.85\columnwidth]{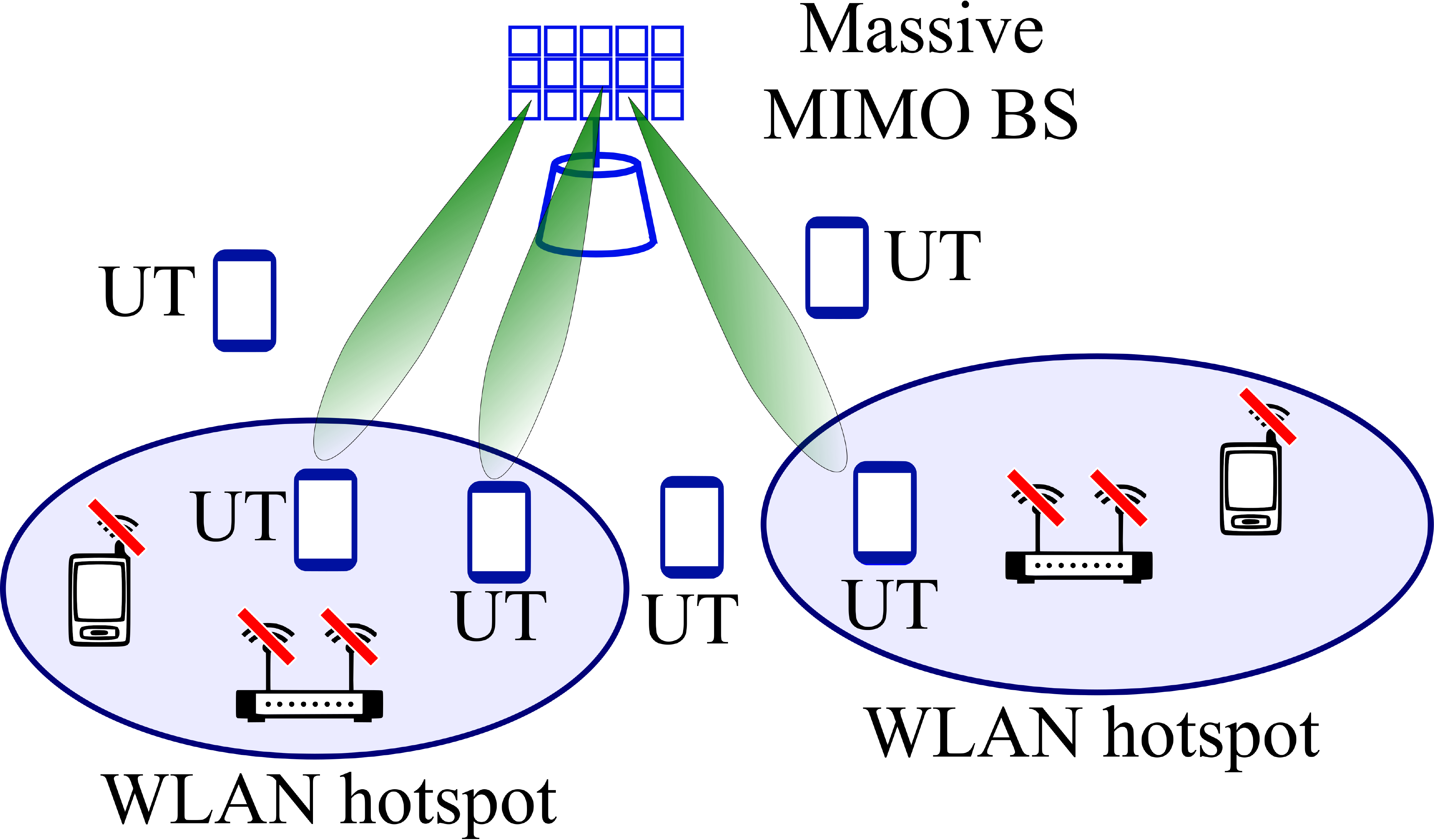}
\caption{UTs in close proximity to a WLAN hotspot served periodically through a conventional LBT phase.}
\label{fig:Scheduling}
\end{figure}

A number of possible initiatives can be adopted to serve UTs located near WLAN hotspots:
\begin{itemize}
\item 
They may be scheduled for transmission in a licensed band, 
if this option is available.
\item
They may be scheduled in the unlicensed band, in a different, less crowded, channel, 
where they might not be affected by WLAN transmissions in their vicinity.
\item 
They may be served in the unlicensed band in the same channel, 
when the channel conditions have changed, 
e.g., because UTs or WLAN devices have moved. 
This option may be viable for non-delay-sensitive applications.
\item 
They may be served in the unlicensed band in the same channel,
by periodically reverting to conventional LBT operations, 
i.e., through discontinuous transmission as shown in Fig.~\ref{fig:Scheduling}. 
Such approach ensures that all UTs can be served in the unlicensed band, 
at the expense of a lower spatial reuse.
\end{itemize}


\subsection*{Data Transmission: Beamforming with Radiation Nulls}

Thanks to the large number of available d.o.f., 
a mMIMO-U BS is able to spatially multiplex the selected UTs in downlink, 
while forcing $N_{\rm N}$ nulls on the channel subspace occupied by the neighboring WLAN devices, 
as depicted in Fig.~\ref{fig:SystemModel}. 
A variant of the well-known zero-forcing precoder can be employed for this purpose~\cite{HoyHosTen2014}.\footnote{
From a mathematical perspective, 
dedicating $N_{\rm N}$ d.o.f. to radiation nulls is equivalent to using a virtual mMIMO array with $N_{\rm N}$ less antennas in highly-scattered environments.}

Due to the limited number of antennas available at the UTs, 
these cannot suppress interference towards or from WLAN devices during mMIMO-U uplink operations, 
e.g., UT-to-BS data or pilots. 
Uplink transmissions are therefore, in general, more challenging than those in the downlink, 
as highlighted in a subsequent section.
\section*{mMIMO-U Use Cases}

In what follows, we present use cases where mMIMO-U can make a difference, and discuss specific examples.

\begin{figure*}[!t]
\centering
\includegraphics[width=\figwidth]{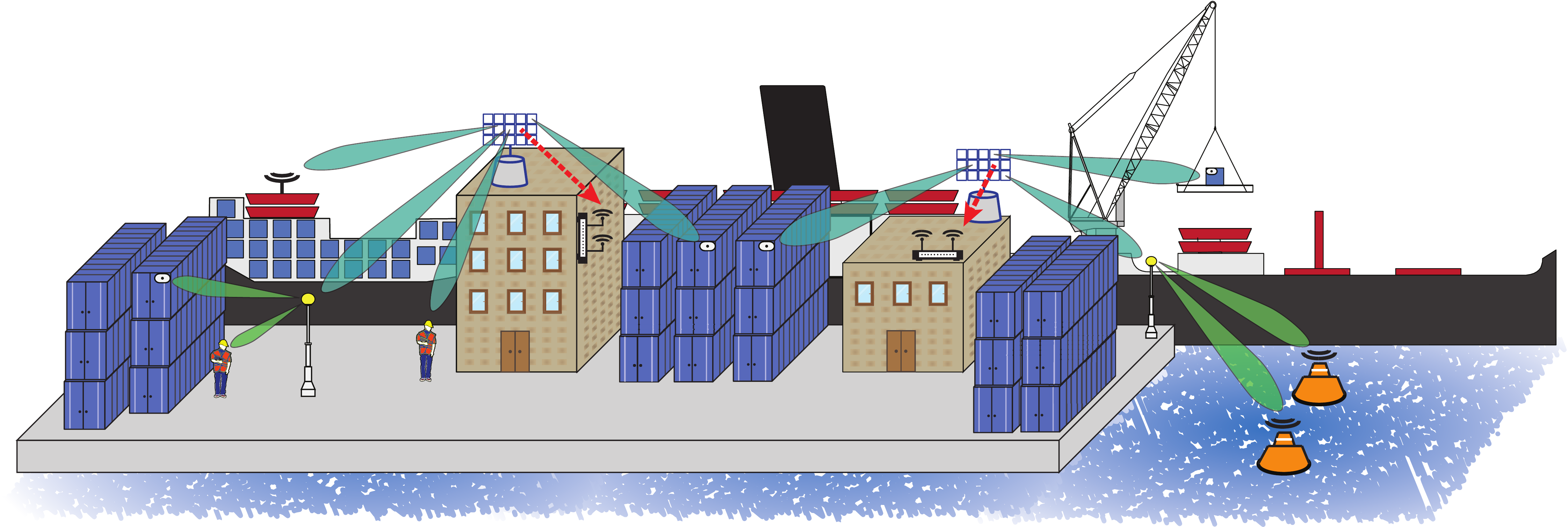}
\caption{Example of a mMIMO-U deployment --- A commercial port with: unlicensed backhaul links to small cells mounted on lamp posts, unlicensed access links to various end users, and radiation nulls to coexisting WLAN devices.}
\label{fig:Port}
\end{figure*}

\subsection*{mMIMO-U for Access}

mMIMO-U can be used to provide direct access to end-users. This specific use case can be divided into four different sub-cases: 
indoor-private, indoor-public, outdoor-private, and outdoor-public access. 

In all combinations where the word \emph{public} is present, 
one can assume that cellular coverage is also available, and thus solutions like LWA and LAA --- which require a licensed anchor --- may be viable. 
An illustrative example of an indoor-public scenario could be a shopping mall, 
while for an outdoor-public setting could think of a stadium. 
In these venues, the adoption of mMIMO-U in LWA and LAA can improve performance and coexistence with WLANs already deployed.

In the case of private venues, both indoor and outdoor, 
owners are generally concerned about sharing sensitive data with cellular operators, 
and thus they prefer to manage and control their own networks. 
Examples of indoor-private and outdoor-private scenarios are corporate buildings and large factories, respectively. 
In this context, MF and WLAN are the two main players, as they do not require a licensed carrier and, as such, do not involve any agreement with an MNO. 
When coupled with MF, 
mMIMO-U can be the way to provide high performance, while addressing the confidentiality concerns of venue owners or public authorities about bringing in third parties to operate their private networks.

\subsection*{mMIMO-U for Backhaul}

The mMIMO-U technology can be also efficiently adopted to provide stable unlicensed wireless backhaul connection to a number of outdoor small cells. 
A first argument to be made in support of this use case is that MNOs are not particularly keen to divide their scarce and valuable licensed spectrum to provide both wireless access and backhaul connections at the same time. 
A second consideration is that unlicensed spectrum is usually less loaded outdoors than indoors, 
as WLAN deployments are mostly within premises.
Finally, having a cost-effective backhauling solution translates into a potential enabler of ultra-dense small cell deployments. 
In order to serve this use case and to provide reasonably large backhaul capacities,
mMIMO-U should be coupled with the use of multiple antennas at the small cells. 

Critical scenarios where we envision the adoption of mMIMO-U for backhaul applications are large commercial areas, 
such as the seaport depicted in Fig.~\ref{fig:Port}. 
In these challenging environments, the logistics and traceability of people and goods is critical, 
and they require a solid and well distributed wireless network with a dense deployment of small cells. 
mMIMO-U is an appealing solution in this case, 
as it provides a cost-effective backhaul through its non-line-of-sight multiplexing capabilities.

\section*{Performance of mMIMO-U for Outdoor Access}

We now evaluate the performance of outdoor mMIMO-U deployments by considering a wrapped-around hexagonal cellular layout with 19 ten-meter high sites, three sectors per site, and $150$~m inter-site distance.
The number of antenna elements considered at the mMIMO-U antenna array are 16, 32, 64, and 128.
24 single antenna UTs are uniformly deployed within each BS sector,
and $N_{\rm U}=8$ of them are selected for transmission in each scheduling interval.   
WLAN hotspots of 10~m radius are also uniformly deployed within each sector,
having one AP and eight STAs each.
APs and STAs are located at 1.5 meter height and have a single antenna. In this study we concentrate on a single 20~MHz channel, and assume that WLANs uniformly distribute their 20~MHz transmissions in the four non-overlapping channels of the U-NII-1 band (5.2~GHz).
Path loss models as specified in 3GPP UMi~\cite{3GPP36814} and 3GPP D2D~\cite{3GPP36843} are used for BS-to-device and device-to-device links, respectively, with lognormal shadowing and distance-dependent Ricean fading \cite{3GPP25996}.

When operating in the unlicensed spectrum, 
the maximum transmit power is strictly regulated and must account for the number of spatial d.o.f. used to provide beamforming gain~\cite{FCC2013}. 
The simulations presented in this article abide the regulations by reducing the radiated power according to the beamforming gain provided to each UT~\cite{FCC1430}, 
yielding a total of $30~\textrm{dBm} - 10~\log_{10}~(N_{\rm A}-N_{\rm N})/N_{\rm U}$ for a 20~MHz channel. We also consider that WLAN devices move at low speeds, 
which facilitates a perfect estimation of their aggregate channel covariance matrices.
The WLAN AP and STA transmit powers are 24~dBm and 18~dBm, respectively \cite{FCC2013}.

\subsection*{Coexistence Enhancements}
	
\begin{figure}[!t]
\subfloat[Coexistence in the unlicensed spectrum as seen by WLAN devices.\label{fig:WiFiInt}]{
  \includegraphics[width=0.9\columnwidth]{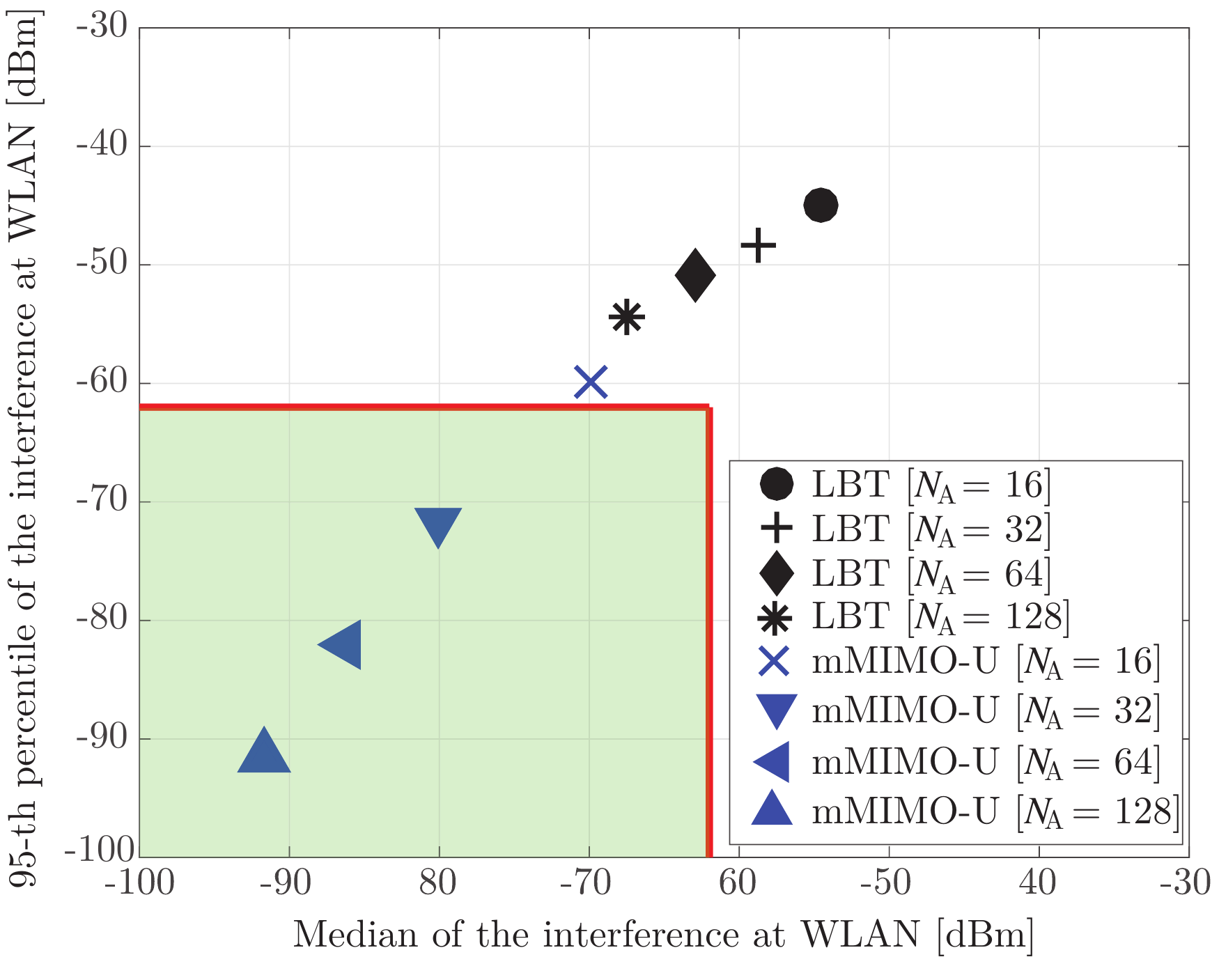}
}\hfill
\subfloat[Coexistence in the unlicensed spectrum as seen by cellular BSs.\label{fig:BSInt}]{
  \includegraphics[width=0.9\columnwidth]{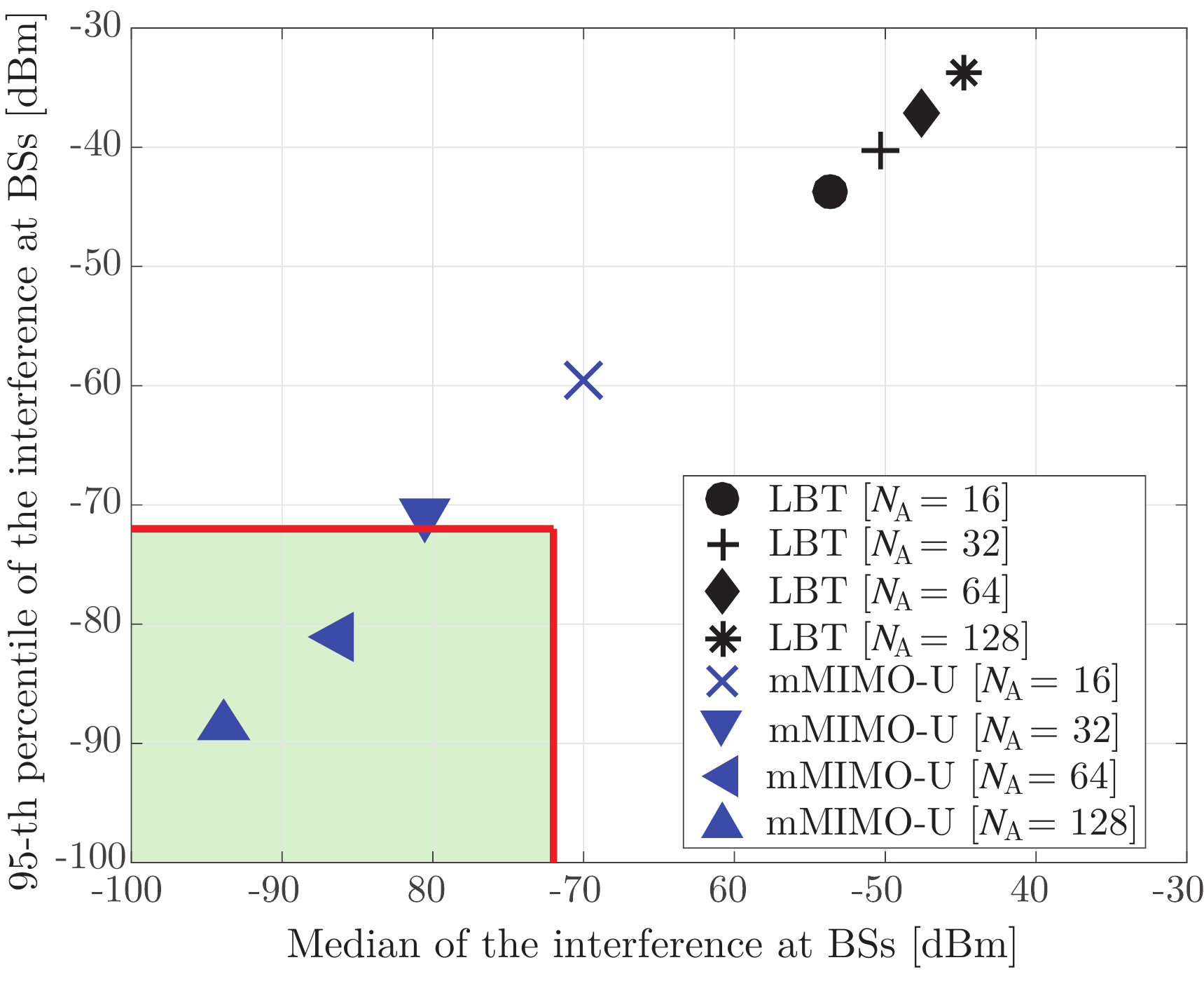}
}
\caption{Coexistence in the unlicensed spectrum in the presence of $2$ active WLAN hotspots per sector as seen by (a) WLAN devices, and (b) cellular BSs.}
\label{fig:coexistence}
\end{figure}

Figs.~\ref{fig:coexistence}(a)~and~\ref{fig:coexistence}(b) show the coexistence amelioration provided by mMIMO-U, 
where $N_{\rm N} = 0.75 \left( N_{\rm A} - N_{\rm U} \right)$ d.o.f. are allocated for nulls, 
with respect to a conventional LBT approach without WLAN interference rejection ($N_{\rm N} = 0$).

In Fig.~\ref{fig:coexistence}(a), 
the perspective of WLAN devices is adopted, 
considering that mMIMO-U BSs have accessed the unlicensed spectrum.
This figure shows the median and 95-th percentile of the aggregate interference received by WLAN devices. 
The green region represents the area where the interfering power received is below the regulatory threshold, $\gamma_{\mathrm{WLAN}} = -62$~dBm~\cite{PerSta2013}, 
and data transmission is feasible.
The results in Fig.~\ref{fig:coexistence}(a) show that 
the aggregate interference at the WLAN devices decreases when the BS is equipped with larger antenna arrays, 
as more d.o.f. are allocated for interference suppression. 
Indeed, both median and 95-th percentile of the aggregate interference fall below $\gamma_{\mathrm{WLAN}}$ for $N_{\rm A}\geq 32$.
In this regime, mMIMO-U enables WLAN devices to access the unlicensed band, 
while BSs are simultaneously transmitting. 
Instead, WLAN devices are not able to commence transmission when BSs do not place radiation nulls, 
since the interference they perceive is above $\gamma_{\mathrm{WLAN}}$.

In Fig.~\ref{fig:coexistence}(b), 
the perspective of cellular BSs is taken, 
assuming that one WLAN device per hotspot has gained access to the unlicensed medium. 
This figure shows the median and 95-th percentile of the aggregate interference perceived by cellular BSs. 
Here, the conservative energy detection threshold value of $\gamma_{\mathrm{BS}} = -72$~dBm is adopted, 
following the specifications of systems such as LAA \cite{3GPP36889}.
In spite of this, Fig.~\ref{fig:coexistence}(b) shows that 
mMIMO-U BSs with a sufficient number of antennas are capable of operating in parallel with WLAN transmissions.
With $N_{\rm A} = 64$ and $N_{\rm A} = 128$ antennas, 
the aggregate interference received by the BSs is $95\%$ of the time smaller than $-81$~dBm and $-89$~dBm, respectively, 
well below the threshold $\gamma_{\mathrm{BS}}$. In contrast, it can observed that BSs without radiation nulls cannot simultaneously share the spectrum with WLAN devices, since their received interference is consistently outside the spatial reuse region (highlighted in green).

\subsection*{Spatial Resource Allocation}

\begin{figure}[!t]
\centering
\includegraphics[width=0.95\columnwidth]{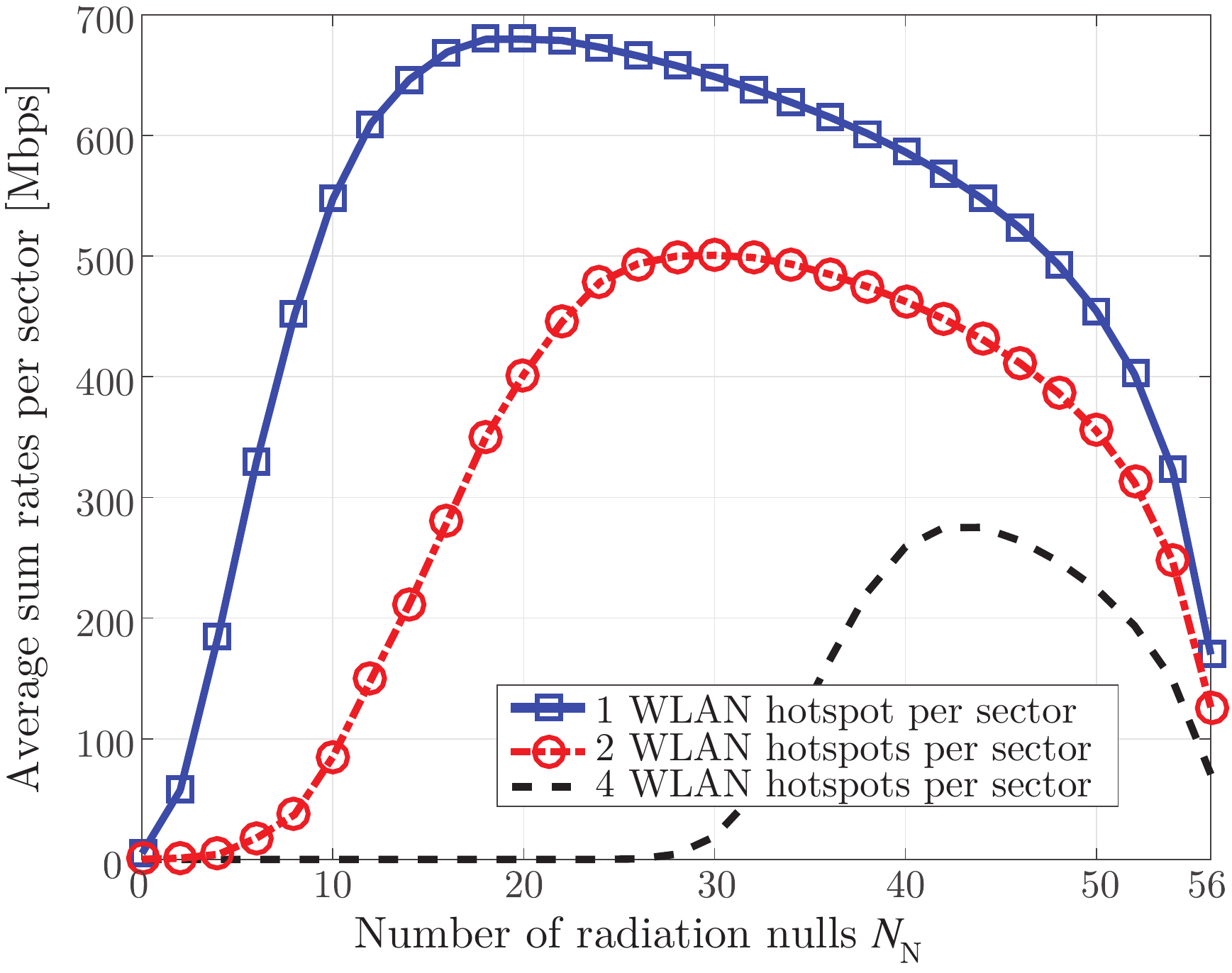}
\caption{Cellular mMIMO-U rates versus number of radiation nulls $N_{\rm N}$ in the presence of $1$, $2$, and $4$ active WLAN hotspots per sector in average.}
\label{fig:Rates_vs_Di}
\end{figure}

In Fig.~\ref{fig:Rates_vs_Di}, we draw the attention to the inherent trade-off between allocating more spatial d.o.f. for WLAN interference suppression and employing them to augment cellular beamforming gain. 
This is illustrated by showing the downlink data rates per cellular sector as a function of $N_{\rm N}$. 
In this figure, $N_{\rm A}=64$ BS antennas 
are considered, 
with full pilot reuse~\cite{Mar:10}. 
The number of radiation nulls $N_{\rm  N}$ allocated for WLAN interference suppression is varied to observe its impact. 
Three scenarios are considered, 
corresponding to $1$, $2$, and $4$ active WLAN hotspots per sector on average.

Importantly, the results of Fig.~\ref{fig:Rates_vs_Di} show that a conventional LBT system with no radiation nulls ($N_{\rm N} = 0$) would not be able to access the channel while WLAN devices are active. 
Instead, mMIMO-U BSs are capable of transmitting as $N_{\rm N}$ increases because the eLBT phase is more likely to be successful. 
However, placing a large number of radiation nulls $N_{\rm N}$ is not recommended once all BSs are able to access the channel, 
since fewer d.o.f. are available for providing multiuser beamforming gains. 
The above trade-off poses a challenge for optimizing the number of nulls $N_{\rm N}$, 
as detailed in the following section. 
Moreover, Fig.~\ref{fig:Rates_vs_Di} shows that coexisting with more WLAN hotspots impacts the attainable cellular rates. 
This is a direct consequence of 
a) the larger interference generated from WLAN devices towards UTs, 
and b) that WLAN devices tend to occupy a larger number of spatial dimensions with increased power, 
which entails placing more radiation nulls for enabling data transmission.

\section*{Challenges}

For mMIMO-U and its promised spectrum reuse to turn into reality, 
several challenges must be first overcome. 
This section is devoted to dissecting the main ones.

\subsection*{WLAN Channel Subspace Estimation}

The spatial awareness of mMIMO-U relies on performing an accurate estimation of the channel subspace occupied by neighboring WLAN devices, 
in order to place radiation nulls and achieve additional spectrum reuse~\cite{GerGarLop2016}. 
In practice, a channel covariance estimate can be obtained by averaging several WLAN symbols received during a silent phase. 
It is obvious that silent phases incur an overhead, 
and an inherent trade-off exists between improving the quality of the covariance estimate and limiting such overhead. 
To reduce the overhead, 
samples acquired during the mandatory eLBT phase can be stored and reused within a validity time. 
A BS can then undergo additional silent phases only when the number of available samples is deemed insufficient. 
Repeated failures of the eLBT phase may indicate that the required number of samples has been underestimated.

\subsection*{Hidden Terminals}

\begin{figure}[!t]
\centering
\includegraphics[width=0.7\columnwidth]{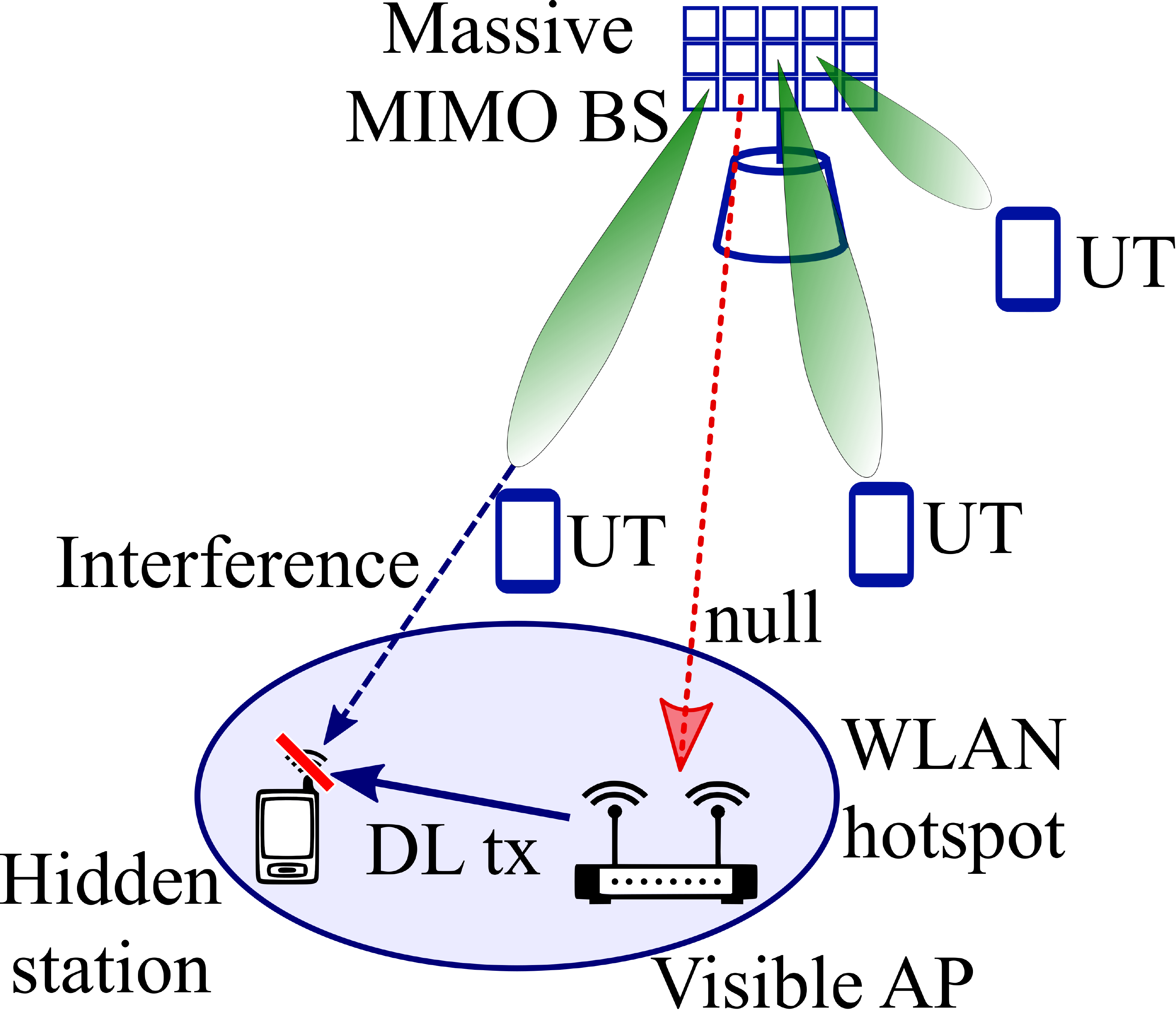}
\caption{Example of hidden terminal problem in mMIMO-U.}
\label{fig:Hidden_Terminal}
\vspace*{-0.3cm}
\end{figure}

As in WLANs, hidden terminal problems may also occur with mMIMO-U operations. 
Consider an AP-to-STA DL-only WLAN transmission, 
as shown in Fig.~\ref{fig:Hidden_Terminal}. 
The lack of traffic from the STA's side might impede the mMIMO-U BS to estimate its channel covariance, 
causing a radiation null to be placed only towards/from the AP.
As a result, the BS might access the channel during the WLAN DL transmission, 
thus disrupting it. 

The above hidden terminal problem highlights the need to perform channel covariance estimation sufficiently often, 
capturing the MAC ACKs or even TCP ACKs, if available, sent by the WLAN STA potentially affected. 
A complementary approach to alleviate this issue consists in periodically reverting to conventional LBT operations with discontinuous transmission. 
A more effective solution may be attained by implementing a network listening mode (NLM) capability at the mMIMO-U. 
NLM allows to decode headers of WLAN packets and to perform a per-device channel covariance estimation, 
maintaining a list of tracked devices. 
Detecting a DL-only WLAN transmission whose recipient is not on the list informs the mMIMO-U BS that one or more WLAN nodes are hidden, 
and that the spectrum should only be accessed through conventional LBT.

\subsection*{Uplink Transmission}

While eLBT and radiation nulls can be used to fully exploit the potential of mMIMO-U in downlink, 
appropriate procedures should be defined for the uplink, 
where collisions between UT-originated pilot or data signals and concurrent WLAN transmissions must be avoided. 

As in conventional mMIMO, 
UT uplink pilots are required at every BS-UT channel coherence interval in order to perform spatial multiplexing. 
In mMIMO-U, a BS may address the scheduled UTs after a successful eLBT with a request to send pilots (RTSP) message. 
RTSP messages should be transmitted with the $N_{\rm N}$ nulls in place, 
such that interference generated at neighboring WLAN devices is suppressed. 
The addressed UTs respond by simultaneously transmitting back omnidirectional pilot signals after a short inter-frame space (SIFS) time interval~\cite{PerSta2013}.\footnote{
UT-WLAN collisions can be minimized by scheduling UTs that are far from WLAN APs. 
UTs may also infer nearby WLAN activities via network allocation vector (NAV) messages and thus defer transmission.}

Similarly to pilot signals, uplink data must be transmitted in a synchronized fashion, 
in order for the BS to perform spatial de-multiplexing. 
This requires BSs to gain access to the medium via an LBT contention phase with no nulls in place, 
and to reserve it through omnidirectional transmissions. 
The channel reservation guarantees that the scheduled UTs can transmit their uplink pilots and data in a conventional multi-user MIMO fashion.

\subsection*{Allocation of Spatial Nulls and Beams}

As shown in Fig.~\ref{fig:Rates_vs_Di}, 
an arbitrary/static assignment of the d.o.f. for radiation nulls can be highly suboptimal, 
especially in dynamic environments. 
To solve this problem, 
a mechanism in which repeated failures of the eLBT phase trigger an increment of the value of $N_{\rm N}$ could be employed. 
Such feedback loop can optimize mMIMO-U performance by adaptively allocating radiation nulls as required, 
thus increasing the probability of performing a successful eLBT.

\section*{Conclusion}

In this article, we have introduced mMIMO-U, a novel solution that could take unlicensed spectrum reuse to a whole new level. With mMIMO-U, intelligent WLAN interference suppression is rewarded with increased transmission opportunities in the unlicensed spectrum. We have identified the fundamental challenges to tackle for this enticing new technology to take off, also paving the way for practical solutions.
We believe that the major improvements attainable over current spectrum sharing approaches make of mMIMO-U a paradigm shift truly worthy of consideration, 
and we envision a plurality of scenarios where it may represent a key enabler.

\vspace*{-0.15cm}

\ifCLASSOPTIONcaptionsoff
  \newpage
\fi
\bibliographystyle{IEEEtran}
\bibliography{Strings_Gio,Bib_Gio}

\begin{thebibliography}{10}
\providecommand{\url}[1]{#1}
\csname url@samestyle\endcsname
\providecommand{\newblock}{\relax}
\providecommand{\bibinfo}[2]{#2}
\providecommand{\BIBentrySTDinterwordspacing}{\spaceskip=0pt\relax}
\providecommand{\BIBentryALTinterwordstretchfactor}{4}
\providecommand{\BIBentryALTinterwordspacing}{\spaceskip=\fontdimen2\font plus
\BIBentryALTinterwordstretchfactor\fontdimen3\font minus
  \fontdimen4\font\relax}
\providecommand{\BIBforeignlanguage}[2]{{%
\expandafter\ifx\csname l@#1\endcsname\relax
\typeout{** WARNING: IEEEtran.bst: No hyphenation pattern has been}%
\typeout{** loaded for the language `#1'. Using the pattern for}%
\typeout{** the default language instead.}%
\else
\language=\csname l@#1\endcsname
\fi
#2}}
\providecommand{\BIBdecl}{\relax}
\BIBdecl

\bibitem{3GPP36360}
{3GPP TS 36.360}, ``Evolved universal terrestrial radio access {(E-UTRA)};
  {LTE-WLAN} aggregation adaptation protocol {(LWAAP)} specification,''
  {v.13.0.0, Mar. 2016}.

\bibitem{3GPP36889}
{3GPP TR 36.889}, ``Feasibility study on licensed-assisted access to unlicensed
  spectrum,'' {v.13.0.0, Jan. 2015}.

\bibitem{jindal2015lte}
N.~Jindal and D.~Breslin, ``{LTE} and {Wi-Fi} in unlicensed spectrum: A
  coexistence study,'' \emph{Google white paper}, June 2015.

\bibitem{MulteFireTechnicalPaper}
{MulteFire Alliance}, ``{MulteFire} release 1.0 technical paper: A new way to
  wireless,'' Jan. 2017.

\bibitem{3GPP-RP-140808}
{3GPP RP 140808}, ``Review of regulatory requirements for unlicensed
  spectrum,'' June 2014.

\bibitem{GerGarLop2016}
G.~Geraci, A.~{Garcia Rodriguez}, D.~L\'{o}pez-P\'{e}rez, A.~Bonfante,
  L.~{Galati Giordano}, and H.~Claussen, ``Operating massive {MIMO} in
  unlicensed bands for enhanced coexistence and spatial reuse,'' to appear in
  \emph{IEEE J. Sel. Areas Commun.}, 2017. Available as arXiv:1612.04775.

\bibitem{Mar:10}
T.~L. Marzetta, ``Noncooperative cellular wireless with unlimited numbers of
  base station antennas,'' \emph{{IEEE} Trans. Wireless Commun.}, vol.~9,
  no.~11, pp. 3590--3600, Nov. 2010.

\bibitem{PerSta2013}
E.~Perahia and R.~Stacey, \emph{Next Generation Wireless {LAN}s: 802.11n and
  802.11ac}.\hskip 1em plus 0.5em minus 0.4em\relax Cambridge University Press,
  June 2013.

\bibitem{HoyHosTen2014}
J.~Hoydis, K.~Hosseini, S.~T. Brink, and M.~Debbah, ``Making smart use of
  excess antennas: Massive {MIMO}, small cells, and {TDD},'' \emph{Bell Labs
  Tech. J.}, vol.~18, no.~2, pp. 5--21, Sept. 2013.

\bibitem{3GPP36300}
{3GPP TR 36.300}, ``Evolved universal terrestrial radio access ({E-UTRA}) and
  evolved universal terrestrial radio access network ({E-UTRAN}); {O}verall
  description,'' {v.14.0.0, Sept. 2016}.

\bibitem{3GPP36814}
{3GPP TR 36.814}, ``Further advancements for {E-UTRA} physical layer aspects,''
  {v.9.0.0, Mar. 2013}.

\bibitem{3GPP36843}
{3GPP TR 36.843}, ``Study on {LTE} device to device proximity services; radio
  aspects,'' {v.12.0.0, Mar. 2014}.

\bibitem{3GPP25996}
{3GPP TR 25.996}, ``Spatial channel model for multiple input multiple output
  ({MIMO}) simulations,'' {v.13.0.0, Dec. 2015}.

\bibitem{FCC2013}
{FCC 662911}, ``Emissions testing of transmitters with multiple outputs in the
  same band,'' Oct. 2013.

\bibitem{FCC1430}
{FCC 14-30}, ``Revision of part 15 of the commission's rules to permit
  unlicensed national information infrastructure {(U-NII)} devices in the 5
  {GHz} band,'' Apr. 2014.

\end{thebibliography}
\end{document}